\documentclass[aps,prl,twocolumn,notitlepage,superscriptaddress,nogroupedaddress]{revtex4-2}

\usepackage{xcolor,graphicx}
\usepackage{amssymb,amsmath,graphicx}
\usepackage[utf8]{inputenc}
\usepackage{comment}
\usepackage{braket}
\usepackage{mathrsfs} 
\usepackage[normalem]{ulem}
\usepackage{hyperref,url}

\begin{document}

\title{Universal defects statistics with strong long-range interactions}

\author{Stefano Gherardini}
\affiliation{CNR-INO, Area Science Park, Basovizza, I-34149 Trieste, Italy}
\affiliation{LENS, Universit\`a di Firenze, I-50019 Sesto Fiorentino, Italy}
\affiliation{Scuola Internazionale Superiore di Studi Avanzati (SISSA), I-34136 Trieste, Italy}

\author{Lorenzo Buffoni}
\affiliation{Department of Physics and Astronomy, University of Florence, 50019 Sesto Fiorentino, Italy}

\author{Nicol\`o Defenu}
\affiliation{Institut f\"{u}r Theoretische Physik, ETH Z\"{u}rich, Wolfgang-Pauli-Str.\,27 Z\"{u}rich, Switzerland}

\begin{abstract}
Quasi-static transformations, or slow quenches, of many-body quantum systems across quantum critical points create topological defects. The Kibble-Zurek mechanism regulates the appearance of defects in a local quantum system through a classical combinatorial process. However, long-range interactions disrupt the conventional Kibble-Zurek scaling and lead to a density of defects that is independent of the rate of the transformation. In this study, we analytically determine the complete full counting statistics of defects generated by slow annealing a strong long-range system across its quantum critical point. We demonstrate that the mechanism of defect generation in long-range systems is a purely quantum process with no classical equivalent. Furthermore, universality is not only observed in the defect density but also in all the moments of the distribution. Our findings can be tested on various experimental platforms, including Rydberg gases and trapped ions.
\end{abstract}

\maketitle

\emph{Introduction:} Universal dynamical scaling is often encountered in many-body systems driven across an equilibrium critical point. Signatures of dynamical universality in classical systems date back to the seminal observations of Kibble and Zurek on the scaling of the density of defects generated by the quasi-static transformation of a many-body Hamiltonian across its critical point~\cite{kibble1976topology,zurek1985cosmological}. Since these early studies, experimental evidences of the Kibble-Zurek mechanism have been outnumbering~\cite{KibblePhysRep80,ZurekPhysRep96,dziarmaga2010dynamics,delcampo2014universality}, making it probably the most accessible example of dynamical universality in modern physics~\cite{kibble2007phase}. Due to quantum-classical correspondence, the Kibble-Zurek mechanism (KZM) shall also apply to slow unitary dynamics across a quantum critical point (QCP). The applicability of KZM to defect scaling at zero temperatures was first conjectured by finite size scaling arguments~\cite{zurek2005dynamics} and, then, demonstrated exactly on the nearest-neighbour Ising model~\cite{dziarmaga2005dynamics}. Several following studies evidenced the universality of the KZM picture and the ubiquity of its applicability, see Ref.~\cite{dziarmaga2010dynamics} for a review.

More recently, KZM  was shown to hold much beyond the traditional result for the defect density and to describe also the scaling of high-order cumulants of the defect statistics~\cite{delCampoPRL18,GomezRuizPRL20,CuiCommPhys20,BandoPRR20}. In fact, as the Ising Hamiltonian is slowly annealed through its ferromagnetic QCP, magnetic domains arise and at the merging points between two domains, a topological defect is formed with a given probability $q$. The formation of each defect is an independent random process over time and, denoting with $L$ the total number of merging points, the probability $p(\ell)$ for the formation of $\ell$ topological defects in spin or fermionic systems is
\begin{equation}\label{eq:binomial_law}
    p(\ell) \sim {\rm B}(\ell,L,q) = \binom{L}{\ell}q^{\ell}(1-q)^{L-\ell}, 
\end{equation}
where $1-q$ is the probability that no defect is originated. In probability theory, this law describes the number of successes in a sequence of independent Bernoulli processes, as e.g.~a coin toss~\cite{vishveshwara2020defect}. While rather remarkable, this phenomenon is in some sense \emph{disappointing}, since the quantum nature of the system and the coherent dynamics appear to leave no trace in the defect distribution, which is identical to the one of a purely classical process. This could be somehow expected since KZM is a consequence of the universal scaling at equilibrium, which may be described in terms of a classical theory in imaginary time~\cite{sachdev1999quantum}.

As we are gonna argue in the following, this classical analogy fails to describe the full counting statistics of defects in the {\it long-range ferromagnetic Ising model}. In fact, while long-range (LR) interactions induce mean-field universal scaling at equilibrium, in the out-of-equilibrium regime they allow to circumvent the constraints imposed by decoherence and equilibration, leading to advantages in quantum technology applications\,\cite{solfanelli2023quantum}. This is particularly true in the strong long-range regime $\alpha<d$, with $d$ the spatial dimension of the system, where the prominent collective character of the long-range systems allowed the observation of a kaleidoscope of novel phenomena, including supersonic propagation of correlations~\cite{jurcevic2014quasiparticle,hauke2013spread,eisert2013breakdown}, fast entanglement spreading and shielding~\cite{richerme2014nonlocal,gong2014persistence,santos2016cooperative,gong2017entanglement,solfanelli2023logarithmic}, anomalous scaling dynamics and ergodicity breaking~\cite{kastner2011diverging,mori2018thermalization,defenu2021metastability}.

\emph{The model:} 
In order to show how long-range interactions promote quantum coherence in the process of defect formation, we are going to consider the full counting statistics of the defects generated by a slow quench across the QCP of the Ising model with flat fully-connected interactions, a.k.a.~the Lipkin-Meshkov-Glick (LMG) Hamiltonian~\cite{lipkin1965validity,meshkov1965validity,glick1965validity}
\begin{align}
\label{lmg}
H=-J\left(\sum_{i<j}\sigma_{i}^{x}\sigma_{j}^{x}+\left(\lambda(t/\tau)+1\right)\sum_{i}\sigma_{i}^{z}\right)
\end{align}
with $t\in [-\tau,\tau]$. The indexes $i,j$ run over all $N$ sites of a one-dimensional lattice. For any slow quench $\tau\to 0$ terminating at or across the QCP, universal scaling of the defects statistics is expected to appear. 

In the following, we will determine the full counting statistics $\mathscr{P}(m)$ of the defects generated by this annealing protocol, demonstrating that $\mathscr{P}(m)$ follows a \emph{negative binomial distribution}. Contrarily to the classical binomial distribution in Eq.~\eqref{eq:binomial_law}, a negative binomial distribution assigns a probability weight to the number of failures $m$ that occur in a sequence of independent Bernoulli trials before $r$ successes are observed. In our case, the event denoted as ``failure" is the formation of a topological defect and occurs with probability $1-s$, with $s$ thus denoting the probability that no defect (``success" event) occurs. The genuinely \emph{quantum nature} of the long-range phenomenology descends from not having an integer value for the number of successes, i.e., $r\in \mathbb{R}$. Hence, the process of defect formation in the quantum long-range Ising model does not have any classical counterpart.

The paper is organized as follows. First, we are going to outline the theory that enables us to describe the crossing of the QCP. Our Hamiltonian is a prototypical model for quantum annealing~\cite{bapst2012quantum} but, despite its integrability, we need to construct an effective low-energy theory to draw an analytical scaling theory. After introducing the analytical solution of our effective theory, we will discuss how our findings breach the KZM and, thus, evade the constraints of adiabatic dynamics. Then, the probability distribution of the internal energy is shown, by determining the analytical expression of the probability to measure the allowed energy values. Remarkably, the probability of obtaining any internal energy of the driven effective model, at a given time of the driving protocol, is identically equal to the full counting statistics of the defects generated within the dynamics. Finally, the consequences of our observations on the quantum thermodynamic behaviour of the LMG Hamiltonian are discussed.

\emph{Effective low-energy theory:} 
The construction of the effective low-energy theory follows the same line as in Ref.~\cite{defenu2018dynamical}. We introduce the leading order Holstein-Primakov approximation~\cite{holstein1940field} of the total spin operator $\boldsymbol{S}=\sum_{i}\boldsymbol{\sigma}_{i}$ with $\boldsymbol{\sigma}_{i}=(\sigma^{x}_{i},\sigma^{y}_{i},\sigma^{z}_{i})$. At equilibrium, this approach reduces the total spin components to position $x$ and momentum $p$ coordinates $S_{z}=N/2-(x^{2}+p^{2}-1)/2$, $S_{x}=\sqrt{N/2}\,x$ and $S_{y}=\sqrt{N/2}\,p$, where operators $x$ and $p$ satisfy the bosonic commutation relation $[x,p]=i\hbar$. It is evident that the aforementioned approximation will remain sensible as long as the system remains close to the full paramagnetic state $\langle S_{z}\rangle/ N \approx -1/2$. This condition is naturally violated during the quasistatic dynamic under consideration since it crosses the equilibrium quantum critical point. However, it is rather straightforward to generalize the procedure in the out-of-equilibrium regime, by aligning the axis of $S_{z}$ with the direction of the instantaneous magnetization, see Ref.~\cite{ruckriegel2011time,lerose2017chaotic}. Assuming that the classical variable describing the instantaneous magnetization adiabatically follows the quasi-static drive, the leading source of deviations from adiabaticity emerges from quantum (energy) fluctuations. The latter are described by the Hamiltonian
\begin{equation}\label{eq:Hamiltonian}
    \mathcal{H}_{t} = \frac{p^2}{2M} + \frac{M}{2}\omega_{t}^{2}x^{2},
\end{equation}
which is obtained from Eq.~\eqref{lmg} by introducing the Holstein-Primakov relations, as shown in the Supplemental Material (SM)~\cite{supp}. In Eq.~\eqref{eq:Hamiltonian}, $M$ is the mass of the oscillator and $\omega_{t}$ denotes the time-dependent driving function. As the system approaches the QCP quantum fluctuations become soft and, accordingly, the frequency follows $\omega_{t} \sim \lambda(t/\tau)$. In order to include effects caused by finite system sizes or small magnetic fields, we will also consider the case of imperfect crossings, i.e. $\lim_{t\rightarrow 0}\omega_{t} = \omega_C \approx 0$ with $t=0$ being the time when the crossing occurs.

In choosing the time-behaviour of $\omega_{t}$, we reproduce the standard Kibble-Zurek protocol~\cite{defenu2021quantum}: at $t=\tau$ the system is prepared in the ground state of $\mathcal{H}_{\tau}$ with a regular and well-separated spectrum ($\omega_{\tau}\gg 0$). Then, the quantum system is driven in order to reduce the spectral gap until the instantaneous spectrum becomes fully degenerate (or nearly so for $\omega_{C}\gtrsim 0$). As customary~\cite{dziarmaga2010dynamics}, we only consider the leading order of the drive in the vicinity of the QCP, so that the time-dependent frequency takes the form $\omega(t)\approx |t/\tau|^{\eta}$, where $\eta$ denotes the critical gap-scaling exponent. Beyond the critical point at $t>0$, the driving function is inverted such that the energy gap opens up again until the dynamic terminates at $t=\tau$.

\emph{Average energy and defect formation:}
The driving function $\omega_{t}$ entails time-dependent variations of the internal energy $\Delta E_t$ of the system. Classically, at any time $t$, the internal energy is a stochastic variable defined as $\Delta E_t \equiv E_{t}^{(m)}-E_{t_0}^{(n)}$, where $\{E^{(n)}\}$ and $\{E^{(m)}\}$ denote the set of eigenvalues of $\mathcal{H}_{t_0}$ and $\mathcal{H}_{t}$ respectively. The latter is formally provided by expressing the system Hamiltonian at $t_0$ and $t$ in spectral representation, i.e., 
\begin{equation}
\mathcal{H}_{t_0} = \sum_{n}E_{t_0}^{(n)}\Pi_{t_0}^{(n)} \quad \text{and} \quad \mathcal{H}_{t} = \sum_{m}E_{t}^{(m)}\Pi_{t}^{(m)},
\end{equation}
where $\Pi_{t_0}^{(n)} \equiv |\phi_{t_0}^{(n)}\rangle\!\langle\phi_{t_0}^{(n)}|$ and $\Pi_{t}^{(m)} \equiv |\phi_{t}^{(m)}\rangle\!\langle\phi_{t}^{(m)}|$ are the projectors on the instantaneous energy eigenstates. Thus, $|\phi_{t'}^{(\ell)}\rangle$ denotes the $\ell$-th eigenfunction of $\mathcal{H}_{t'}$, with $\ell \in \{n,m\}$ unbounded (from above) integer number. Instead, the instantaneous energy values $E_{t}^{(\ell)}$ are equal to $E_{t}^{(\ell)} \equiv \hbar\,\omega_{t}(\ell+\frac{1}{2})$, with $\hbar$ reduced Planck constant.

The corresponding average internal energy $\langle\Delta E_t\rangle$, the expected value of the distribution of $\Delta E_t$, can be interpreted as the average \emph{reversible} quantum (stochastic) work $\langle W\rangle$, provided that the driving function is an \emph{adiabatic} transformation~\cite{DeffnerPRE08}. This means that, in case the drive was not adiabatic, part of the internal energy $\Delta E$ cannot be converted in useful work. Hence, from the first law of thermodynamics $\langle\Delta E_t\rangle = \langle W\rangle_{\rm rev} + \langle W\rangle_{\rm irr}$, in our setting $\langle W\rangle_{\rm irr}=0$ for $t<0$. Instead, the dynamic of the system becomes irreversible from crossing the QCP, and the amount of irreversible work decreases for higher values of $\omega_C$. 
In fact, the average internal energy $\Delta E_t$ of the system can be decomposed as 
\begin{equation}\label{eq:mean_exc_number}
    \langle\Delta E_t\rangle = \frac{\hbar}{2}(\omega_{t}-\omega_{t_0}) + \hbar\omega_{t}\langle \nu_t\rangle \,,
\end{equation}
where the first term on the r.h.s.~represents the adiabatic correction $\langle W\rangle_{\rm rev}$ to the quantum work, while the second term is the non-adiabatic contribution, i.e., the irreversible work $\langle W\rangle_{\rm irr}$. The latter is proportional to the average number of excitations (or defect density) $\langle \nu_t\rangle=\sum_{m}\langle \psi_{t}|m\, \Pi^{(m)}_{t}|\psi_{t}\rangle$ generated by the drive, with $|\psi_{t}\rangle$ the exact time-dependent state of the system.

\begin{figure}[ht!]
\includegraphics[width=0.98\linewidth]{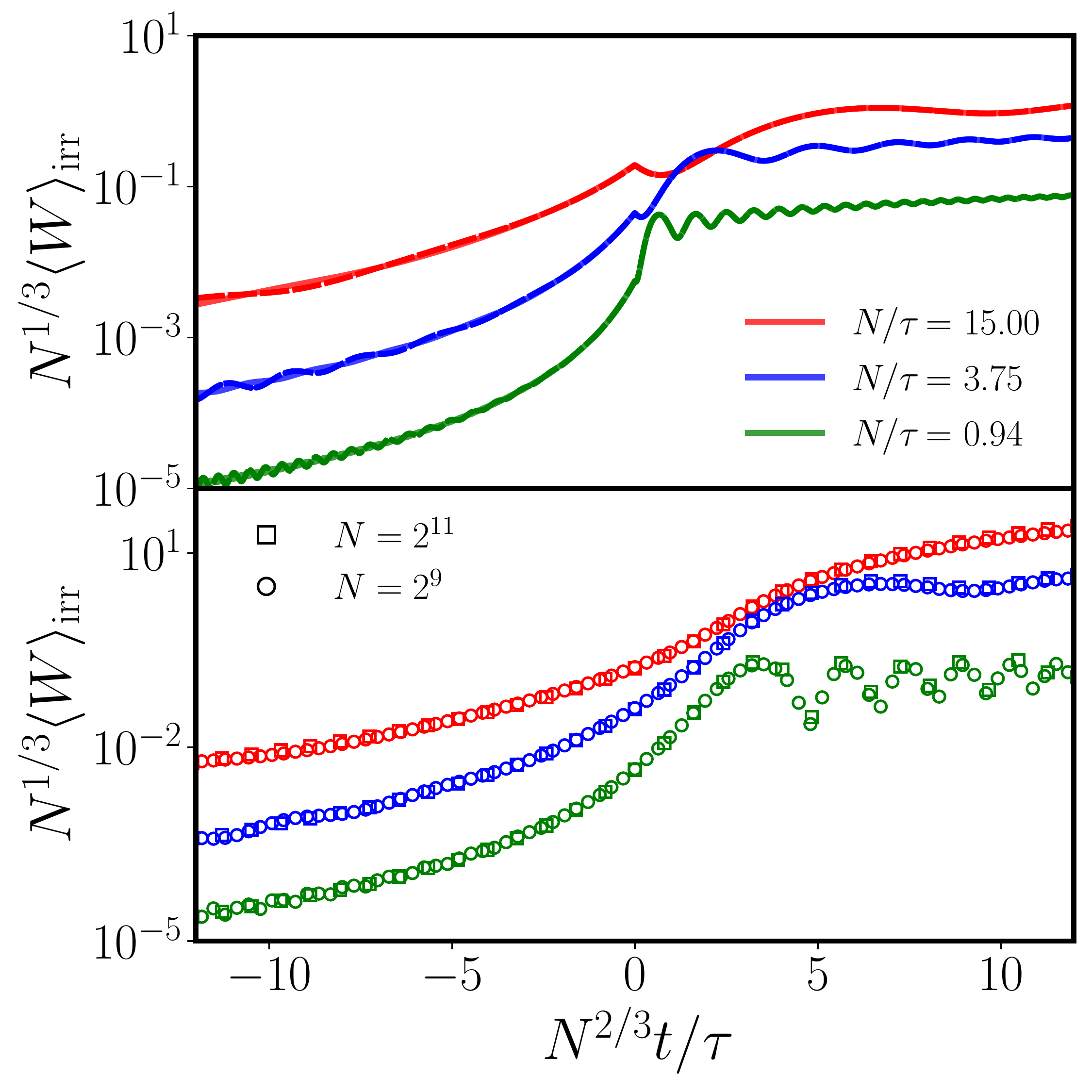}
\caption{\label{Fig1} 
Irreversible work associated respectively to the effective model in Eq.~\eqref{eq:Hamiltonian} (upper panel) and to the full numerical solution of the time-dependent LMG model in Eq.~\eqref{lmg}, with time-dependent coupling $\lambda=t/\tau$, performed in Ref.~\cite{acevedo2014new} (lower panel). Each colour represents a different value of $N/\tau$ (see legend), with $N$ and $\tau$ denoting the system size and drive rate respectively. Dashed and solid lines in the upper panel correspond to two system sizes, $N=2^{9}$ and $N=2^{11}$, which are represented by different symbols for the numerical data (lower panel, see legend). Both the work and the time variables have been rescaled as described in Refs.~\cite{acevedo2014new,defenu2018dynamical}. The similarity between the theoretical model and the numerics demonstrates the capability of the effective model in Eq.~\eqref{eq:Hamiltonian} to describe the universal slow-drive dynamics of the LMG model in Eq.~\eqref{lmg}.}
\end{figure}

Early numerical studies in Ref.~\cite{acevedo2014new} evidenced the lack of any Kibble-Zurek mechanism for the defect density $\langle\nu_t\rangle$ of the LMG model that is quasi-statically driven across its QCP. As a matter of fact, the defect density shows universal behaviour as a function of the combined variable $N/\tau$, where $N$ is the system size and $\tau$ is the drive rate. By choosing $\omega_{C}\propto N^{-1/3}$ and $\eta=1$, our effective model quite perfectly reproduces these exact numerical findings, see Fig.~\ref{Fig1} where we compare the irreversible work obtained by the effective model in Eq.~\eqref{eq:Hamiltonian} (upper panel) and the one found numerically in Ref.~\cite{acevedo2014new} for the LMG Hamiltonian~\eqref{lmg} (lower panel).

\emph{Full counting statistics of defect formation:}
As it was recently shown for the nearest-neighbour Ising Hamiltonian, universality is not only observed in the scaling of the defect density but in the entire full counting statistics of the defects generated by the quasi-static drive~\cite{GomezRuizPRL20}. Within our model, it is also possible to derive the exact expression for the probability to generate a state with energy $E_{t}^{(m)}$, which reads 
\begin{equation}\label{eq:final_prob_excitations}
    p(E_{t}^{(m)}) = \frac{(m-1)!!}{m!!}\sqrt{1-|R_t|^{2}}\,|R_t|^{m}\,,
\end{equation}
where $m\in 2\mathbb{N}$ and
\begin{equation}\label{eq:modulus_Rt_square}
    |R_t|^{2} \equiv \frac{\left(\frac{1}{2\xi_{t}^{2}}-\omega_{t}\right)^{2}+\frac{\dot{\xi}_{t}^{2}}{\xi_{t}^{2}}}{\left(\frac{1}{2\xi_{t}^{2}}+\omega_{t}\right)^{2}+\frac{\dot{\xi}_{t}^{2}}{\xi_{t}^{2}}}\,.
\end{equation}
In Eq.~(\ref{eq:modulus_Rt_square}), $\xi_{t}$ denotes the time-dependent length of the oscillator, which obeys the Ermakov equation as discussed in the SM~\cite{supp}. From Eq.~\eqref{eq:final_prob_excitations} one can derive the exact expression for all correlations of the defect density in terms of the coefficient $|R_{t}|^{2}$~\cite{supp}. In particular, notice that
\begin{equation}
\langle \nu_t\rangle = \frac{ |R_t|^{2} }{ 1-|R_t|^{2} } \quad \text{and} \quad \langle \nu_t^{2}\rangle = \frac{ |R_t|^{2}(2+|R_t|^{2}) }{ (1-|R_t|^{2})^{2} }
\end{equation}
that exclusively depend on $\omega_t$ and $\xi_t$ via $|R_{t}|^{2}$. Hence, all the cumulants of the full counting statistics of the defects obey the same scaling theory observed for the average defect density in Fig.~\ref{Fig1}.
This scaling theory contradicts the standard Kibble-Zurek expectation since the mechanism of defect formation is independent of the drive rate $\tau$ in the large $\tau$ limit. Rather, in the case of perfect degeneracy, the only parameter influencing the defect formation is the critical gap-scaling exponent $\eta$.

\begin{figure*}[ht]
    \centering
    \includegraphics[width=1\linewidth]{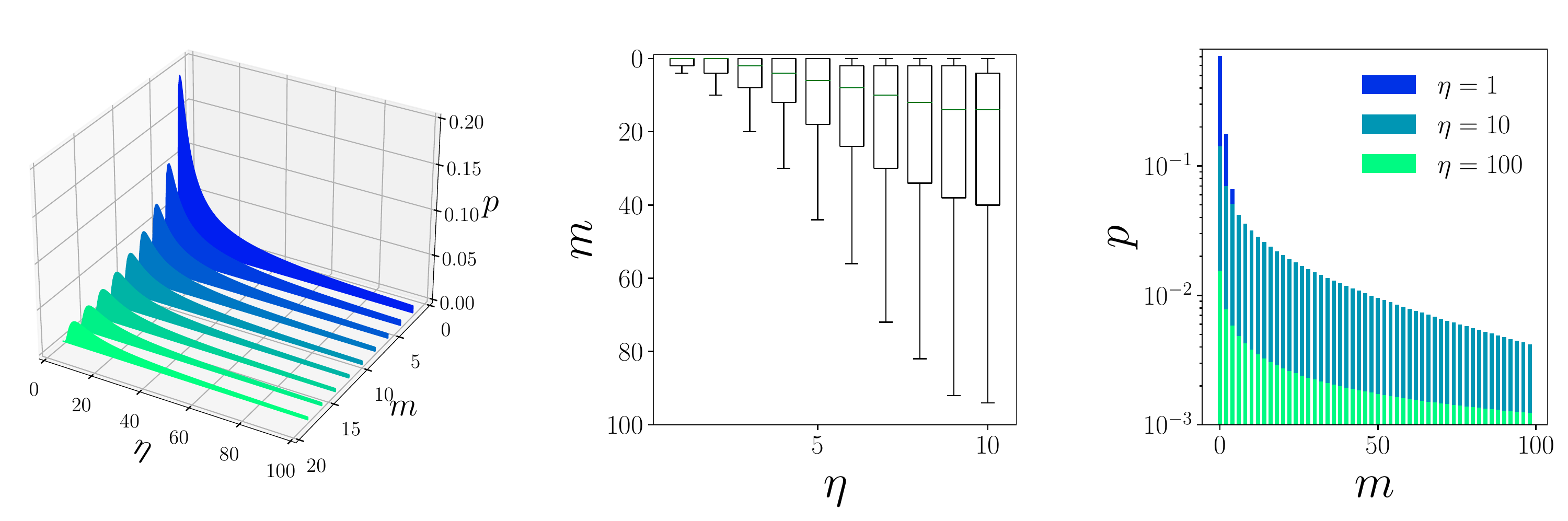}
    \caption{
    Left: Plot of the probability $p(E_{t}^{(m)}) = p(\nu_{t} = m)$ in Eq.~\ref{eq:final_prob_excitations}, negative binomial distribution, as a function of $\eta\in[0,100]$ ($m\in 2\mathbb{R}$) for different (even) values of $m\in [2,18]$ in the limit $\tau \to \infty$. Mid: Top-down view of the distributions at fixed $\eta$. They are represented as a box plot with average value (green solid line), standard deviation (rectangular boxes) and domain (lower and upper extremes). Right: Probability distributions at $\eta = 1$, $\eta=10$ and $\eta = 100$ respectively (in log scale). We can observe how the number of energy excitations, which are peaked around zero for small values of $\eta$, quickly spread out to higher values of $m$ as $\eta$ increases. 
    }
    \label{Fig2}
\end{figure*}

This result is evident by explicitly considering the perfect quasi-static limit $\tau\to\infty$ in such a way that the energy probability distribution coincides with the probability of defect formation after a perfect quantum annealing protocol: $\lim_{\tau\to\infty}p(E_{\tau}^{(m)}) \equiv \mathscr{P}(m)$. Interestingly, in this limit, one has an analytic expression for $|R_{t}|$, which reads
\begin{align}
\lim_{\tau\to\infty}|R_{\tau}|=\cos\left(\frac{\pi}{2+\eta}\right).
\end{align}
Thus, inserting the latter expression in Eq.~\eqref{eq:final_prob_excitations}, one obtains the exact expression of the probability for defect formation in a quantum long-range ferromagnet after a slow quantum annealing procedure, i.e.,
\begin{align}
\label{fin_dist}
\mathscr{P}(m)=\binom{m-1/2}{m}\sin\left(\frac{\pi}{2+\eta}\right)\cos\left(\frac{\pi}{2+\eta}\right)^{m}.
\end{align}  
which is a negative binomial distribution of fractional index $r=1/2$~\cite{degroot1986probability}. Due to the fractional index, the process of defect formation in long-range quantum systems is a purely quantum process and cannot be related to a classical extraction problem as it occurs in the defect formation of local quantum systems~\cite{vishveshwara2020defect}. This finding is the main result of the paper.

Remarkably, Eq.~\eqref{fin_dist} also entails the \emph{universality} of the distribution of the internal energy, as well as of the distributions of both the number of excitations and the formation of topological defects in the quasi-static limit~\cite{supp}. Of course, the universality of such distributions is reflected in the universality of the corresponding statistical moments, which only depend on the critical gap-scaling exponent $\eta$. The probability distribution is plotted in Fig.~\ref{Fig2} as a function of both $m \in 2\mathbb{N}$ (number of defects) and $\eta$. Greater is the value of $\eta$, and larger is the support of $\mathscr{P}(m)$ along the $m$-axis. However, both the average value and the variance of $\nu_{t}$, which have to take a finite value due to the properties of the negative binomial distribution, tend to a constant stationary value as $\eta$ increases, see the central panel in Fig.~\ref{Fig2}. The difference with local and fermionic systems \cite{delCampoPRL18,GomezRuizPRL20,CuiCommPhys20,BandoPRR20} is thus evident, since defect formation in long-range systems is strongly affected by the infinitely-degeneracy of the quasi-particle spectrum.

\begin{figure}
    \centering
    \includegraphics[width=0.98\linewidth]{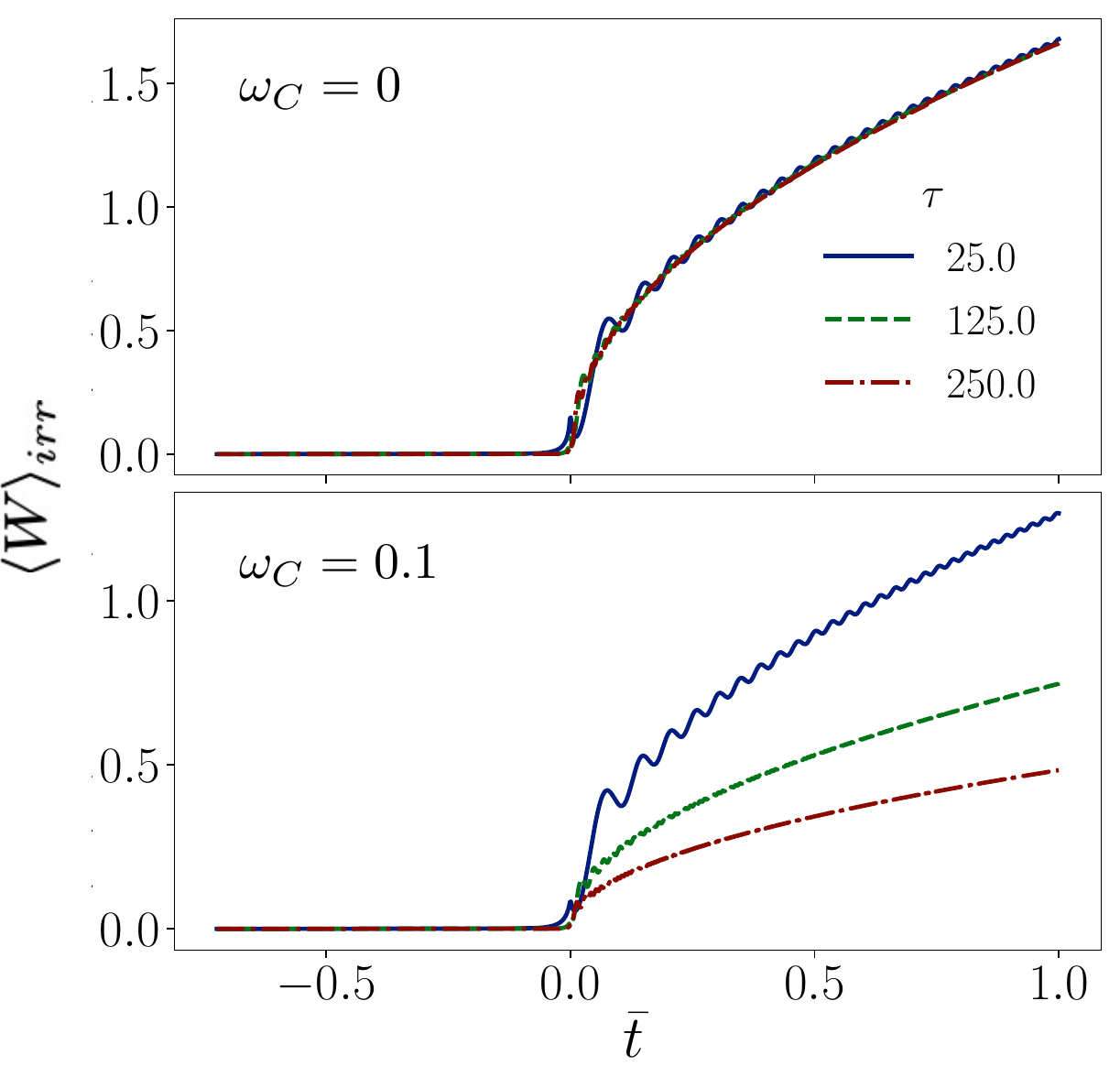}
    \caption{
    Average irreversible work $\langle W\rangle_{\rm irr}$ as a function of the normalized time $\bar{t}=t/\tau$ for different values of $\tau$. In the top panel we can observe that, for $\omega_C \approx 0$, increasing $\tau$ does not help restore the adiabaticity of the system, and thus also $\langle W\rangle_{\rm irr}$ cannot be decreased. Instead, in the bottom panel, for a small (but finite) $\omega_C = 0.1$ (in dimensionless units), it is possible to reduce the effect of irreversibility by increasing the time $\tau$, as predicted by the quantum adiabatic theorem~\cite{landau1969statistical,kato1950adiabatic}.}
    \label{Fig3}
\end{figure}

\emph{Quantum thermodynamics at criticality:} 
Let us now consider the thermodynamic interpretation. First, while crossing the QCP at $t=0$, a significant irreversible component of work is generated on average. In fact, as shown in Fig.~\ref{Fig3}, before $t=0$ the mean internal energy $\langle\Delta E_t\rangle$ perfectly follows the time-dependence imposed by the driving function $\omega_t=|t|$, with $\eta=1$ in the figure, thus entailing $\langle W\rangle_{\rm irr}=0$ for $t<0$. Conversely, crossing the QCP makes irreversible the quantum system dynamics, and such irreversibility gets more pronounced as $\omega_C$ decreases, with $\omega_C$ quantifying the distance from perfect criticality. 
For $\omega_C$ approaching zero, irreversible work is generated not only on average but also at the level of fluctuations. This is evident from the fact that the variance of $\Delta E_t$ is proportional to ${\rm Var}(\nu_t)$ at any time $t$. From the distinction between reversible and irreversible work in Eq.~(\ref{eq:mean_exc_number}) and Fig.~\ref{Fig3}, ${\rm Var}(\Delta E_t)$ has to be interpreted as the variance of the irreversible work for $\omega_C \rightarrow 0$ (numerical evaluations of ${\rm Var}(\Delta E_t)$ are in the SM~\cite{supp}).

Fig.~\ref{Fig3} demonstrates the irreversibility of the work process when the QCP is perfectly achieved for $\omega_{C}=0$. In general, the \emph{quantum adiabatic theorem}~\cite{landau1969statistical,kato1950adiabatic} guarantees that on average the internal energy of a driven quantum system tends to vanish in the limit of slow driving, i.e., $\tau\rightarrow\infty$. For any $\omega_{C}>0$, indeed, the internal energy of the system vanishes with a scaling depending on $\tau$~\cite{DeffnerPRE08}. Yet, the statements of the quantum adiabatic theorem only hold for a non-degenerate spectrum, i.e., $\omega_{C} \neq 0$. In fact, for $\omega_{C} \neq 0$ ($\omega_{C} = 0.1$ in the upper panel of Fig.~\ref{Fig3}), longer is the protocol duration $\tau$ and smaller is $\langle W\rangle_{\rm irr}$. Conversely, if $\omega_{C} \to 0$, then $\langle W\rangle_{\rm irr}/(\hbar\omega_t) = \langle\nu_t\rangle$ is universal and independent of $\tau$. As a result, the theses of the quantum adiabatic theorem are never recovered, neither at $\tau\rightarrow\infty$, meaning that the driving process at the quantum criticality occurs irreversibly. 

\emph{Conclusions:}
In this paper, we answer the question about how the statistics of topological defects formation, the breaking of the quantum adiabatic theorem and the production of energy excitations are linked to each other in quantum long-range systems. We have here determined that, contrarily to local quantum systems~\cite{delCampoPRL18,GomezRuizPRL20,CuiCommPhys20,BandoPRR20}, the finite-time crossing of an infinitely-degenerate quantum critical point gives rise to energy excitations in the system dynamics (aka topological defects formation) that in probability follows a negative binomial distribution, with a fractional number ($r=\frac{1}{2}$) of `success' events (no defects). This kind of probability law has no analogous counterpart in classical combinatorial problems.

These concepts are connected with the first law of thermodynamics in driven quantum systems crossing an infinitely-degenerate critical point. We have thus shown how to distinguish among the distributions of reversible and irreversible work, by providing an explanation in energetic terms of the corresponding adiabatic and non-adiabatic regimes. The latter are originated by the breaking of the quantum adiabatic theorem due to the spectral degeneracy at $\omega_{C}=0$. In fact, until the quantum critical point is not reached, the system can be driven adiabatically and no irreversible work is generated. Then, in correspondence of the criticality, any statistical moments of the internal energy distribution become non-analytic, and immediately after, in the non-adiabatic regime, the dynamics become irreversible for a period that is inversely proportional to $\omega_C$, thus diverging for $\omega_C=0$. This kind of irreversibility generates energy excitations. Such a feature is universal, and therefore it is expected to find application in several long-range quantum systems and, in particular, in those with flat interactions~\cite{acevedo2014new}.

Our predictions shall find experimental observation in several quantum platforms, which are capable of realizing fully-connected interactions and complex geometries. In particular, currently, fully-connected interactions can  be realized in trapped ion system, which allow to also implement power-law decay interactions with tunable exponents in isolated environments~\cite{monroe2021programmable}. Correspondingly, cold atoms trapped in optical cavities have been traditionally employed to engineer flat interacting systems~\cite{mivehvar2021cavity}, which can be mapped into spin Hamiltonians such as Eq.~\eqref{lmg}~\cite{defenu2023long}. In such systems, the process of defect formation has been shown to receive a substantial contribution from dissipation~\cite{brennecke2013real}. Finally, Rydberg atom systems---already used to investigate defect formation in quantum systems~\cite{keesling2019quantum,chomaz2022dipolar}---are a promising platform to observe our predictions, as they have been recently employed to realize tunable complex geometries, thus to tailor customary interaction patterns~\cite{periwal2021programmable}.

\begin{acknowledgments}
\emph{Acknowledgments:} 
S.G. warmly thanks Ricardo Puebla for discussions, especially about irreversible work generation from crossing the QCP in the case study here considered. 
N.D. acknowledges useful discussion with G. M. Graf during the early stages of this work.
This work was supported by The Blanceflor Foundation for financial support through the project ``The theRmodynamics behInd thE meaSuremenT postulate of quantum mEchanics (TRIESTE)'' [S.G. and N.D.], the European Commission under GA n.~101070546--MUQUABIS, the PNRR MUR project PE0000023-NQSTI [S.G.] and the PNRR MUR project SOE0000098 [L.B.].  This research was funded in part by the Swiss National Science Foundation (SNSF) [200021\_207537]. The support of the Deutsche Forschungsgemeinschaft (DFG, German Research Foundation) under Germany’s Excellence Strategy EXC2181/1-390900948 (the Heidelberg STRUCTURES Excellence Cluster) is also acknowledged [N.D.].
\end{acknowledgments}

\clearpage
\newpage

\onecolumngrid

\section{Supplemental Material}

\subsection{Holstein-Primakoff expansion}

We here review the mapping between the Lipkin-Meshkov-Glick (LMG) Hamiltonian, see Eq.~(2) of the main text, and the Hamiltonian of a driven quantum harmonic oscillator, Eq.~(3) of the main text, at the lowest order in the $1/N$ expansion. The procedure is analogue to the one described in~\cite{dusuel2005continuous}. 

First, we rewrite the LMG Hamiltonian as
\begin{align}
\label{lmg_h}
H=-J\left(\frac{1}{N}\sum_{i,j;\,\,i<j}\sigma^{x}_{i}\sigma^{x}_{j}+h\sum_{i}\sigma^{z}_{i}\right)=-2J\left(\frac{1}{N}S_{x}^{2}+h\,S_{z}\right)+\frac{J}{2} \,,
\end{align}
where ${\bf S} \equiv (S_{x},S_{y},S_{z}) = \sum_{i}{\bf \sigma}_i$ is the total spin operator as defined in the main text. Without loss of generality, we will set $J=1$ in the following.

Secondly, the coordinate frame of the total spin operator is spatially rotated via the rotation matrix (unitary operator)
\begin{align}
R=\begin{pmatrix}
\cos\theta\cos\phi & \cos\theta\sin\phi & -\sin\theta\\
-\sin\phi & \cos\phi & 0\\
\sin\theta\cos\phi & \sin\theta\sin\phi & \cos\theta
\end{pmatrix}.
\end{align}
The rotation matrix $R$ gives rise to the transformation $\boldsymbol{\Sigma}=R\boldsymbol{S}$, where $\boldsymbol{\Sigma}$ is the rotated total spin operator, such that
\begin{align}\label{sub_rule}
\boldsymbol{S}=R^{T}\boldsymbol{\Sigma}=\begin{pmatrix}
\cos\theta\cos\phi & -\sin\phi  &\sin\theta\cos\phi \\
\cos\theta\sin\phi & \cos\phi & \sin\theta\sin\phi\\
-\sin\theta &   0 & \cos\theta
\end{pmatrix}\begin{pmatrix}
\Sigma_{x}\\
\Sigma_{y}\\
\Sigma_{z}
\end{pmatrix},
\end{align}
with $R^{T}= R^{-1}$ defining the inverse transformation. From Eq.~(\ref{sub_rule}), we can get the expression of the components $S_x$, $S_y$, $S_z$ of the total spin operator as a function of both the angles $\theta$ and $\phi$ and the components of $\boldsymbol{\Sigma}$. In this way, the LMG Hamiltonian in the rotated frame is equal to
\begin{align}\label{eq:rotated_H_LMG}
H=-\frac{2}{N}\Big(\cos\theta\cos\phi \,\Sigma_{x}-\sin\phi\,\Sigma_{y}+\sin\theta\cos\phi\,\Sigma_{z}\Big)^{2} - 2h\Big(-\sin\theta\,\Sigma_{x}+\cos\theta\,\Sigma_{z}\Big)+\frac{1}{2} \,.
\end{align}

Thirdly, the rotated coordinate frame is aligned with the equilibrium magnetization frame of the model. This is achieved by choosing
\begin{subequations}
\begin{align}
\label{mgn_alignment}
\theta&=\begin{cases}
0 & \text{if $h>h_{c}$}\,,\\
\arccos(h) & \text{if $h<h_{c}$}\,,
\end{cases}\\
\phi&=0\,,\label{mgn_alignment2}
\end{align}
\end{subequations}
where $h_{c}$ is the location of the quantum critical point.
It is worth noting that, by substituting \eqref{mgn_alignment} and \eqref{mgn_alignment2} into \eqref{sub_rule}, we recover the rotation matrix introduced in Eq.~(27) of Ref.~\cite{dusuel2005continuous}.

We are now in the position to introduce the Holstein-Primakoff expansion, which shall be valid around the ground state of the LMG Hamiltonian. The expansion consists in expressing the $N$-component rotated spin operator $\boldsymbol{\Sigma}$ in terms of the variables of the quantum harmonic oscillator, i.e.,
\begin{subequations}
\begin{align}
\frac{\Sigma_{x}}{N}&=\frac{x}{\sqrt{2N}}+O\left(N^{-3/2}\right)=\frac{1}{2\sqrt{N}}\left(a^{\dagger}+a\right)+O\left(N^{-3/2}\right),\label{eq_Primakoff_1}\\
\frac{\Sigma_{y}}{N}&=\frac{p}{\sqrt{2N}}+O\left(N^{-3/2}\right)=\frac{i}{2\sqrt{N}}\left(a^{\dagger}-a\right)+O\left(N^{-3/2}\right),\label{eq_Primakoff_2}\\
\frac{\Sigma_{z}}{N}&=\frac{1}{2}-\frac{x^{2}+p^{2}-1}{2N}+O\left(N^{-2}\right)=\frac{1}{2}-\frac{a^{\dagger}a}{N}+O\left(N^{-2}\right),\label{eq_Primakoff_3}
\end{align}
\end{subequations}
where $x$, $p$ are fictitious position and momentum operators of an effective quantum harmonic oscillator that describes the action of the rotated total spin operator in the proximity of the LMG ground state. Moreover, $a^{\dagger}$ and $a$ are respectively the creation and annihilation bosonic operators. Notice that the validity of the Holstein-Primakoff expansion improves as the value of $N$ increases. In this way, using (\ref{eq_Primakoff_1}), (\ref{eq_Primakoff_2}), (\ref{eq_Primakoff_3}) to represent the LMG Hamiltonian (\ref{eq:rotated_H_LMG}), we obtain the following $1/N$-expansion for $H$:  
\begin{align}
\frac{H}{N}=-\sum_{\ell=0}^{2}\frac{e_{\ell}}{N^{\ell/2}}
\end{align}
with coefficients
\begin{subequations}
\begin{align}
e_{0}&=\frac{\sin\theta^{2}}{2}+h\cos\theta \,,\\
e_{1}&=\frac{\left(\sin(2\theta)-2h\sin\theta\right)}{\sqrt{2}}x \,,\\
e_{2}&=-\left(\sin^{2}\theta+h\cos\theta-\cos^{2}\theta\right)x^{2}-(\sin^{2}\theta+h\cos\theta)p^{2}+\sin^{2}\theta+h\cos\theta-\frac{1}{2}\,.
\end{align}
\end{subequations}
Notice that, under this approximation, all the higher-order terms in $1/N$ are neglected. Of course, being the quantities $e_{\ell}$ function of the values of the angles $\theta$ and $\phi$ through Eq.~\eqref{mgn_alignment}, their expressions change depending on the phase diagram region where the LMG Hamiltonian is analyzed. Specifically, from Eq.~\eqref{mgn_alignment}, the zeroth-order term $e_{0}(h)$ is equal to
\begin{align}
e_{0}(h)=\begin{cases} 
h & \text{if $h>h_{c}$}\,,\\
\frac{1+h^{2}}{2} & \text{if $h<h_{c}$}\,,
\end{cases}
\end{align}
in agreement with the result in Ref.~\cite{dusuel2005continuous}. Conversely, it can be shown that $e_1$ is vanishing in both the symmetry-broken and the symmetric phases, due to the alignment of the rotated coordinate frame with the equilibrium magnetization axes. As a result, the leading-order contribution of the LMG Hamiltonian to quantum fluctuations around the ground state, in term of the Holstein-Primakoff expansion, is 
\begin{align}\label{eq:leading_order_harmonic_osc}
-e_{2}=\frac{p^{2}}{2m}+\frac{m}{2}\omega^{2}x^{2}-\delta e'\,,
\end{align}
where
\begin{align}
m=\begin{cases} 
\frac{1}{2h} & \text{if $h>h_{c}$}\,,\\
1/2 & \text{if $h<h_{c}$}\,,
\end{cases}
\qquad
\omega^{2}=\begin{cases} 
4h(h-1) & \text{if $h>h_{c}$}\,,\\
4(1-h^{2}) & \text{if $h<h_{c}$}\,,
\end{cases}
\qquad
\delta e'=\begin{cases} 
h-\frac{1}{2} & \text{if $h>h_{c}$}\,,\\
\frac{1}{2} & \text{if $h<h_{c}$}\,.
\end{cases}
\end{align}
We can note that the right-hand-side of Eq.~(\ref{eq:leading_order_harmonic_osc}) is ascribable to a harmonic oscillator Hamiltonian, which thus can be diagonalized in terms of the creation and annihilation operators
\begin{align}\label{lab_hh_diag}
b=\sqrt{\frac{m\omega}{2}}\left(x + \frac{i}{m\omega}p\right),\qquad
b^{\dagger}&=\sqrt{\frac{m\omega}{2}}\left(x - \frac{i}{m\omega}p\right).
\end{align}
In this way,
\begin{align}
\label{ld_corr}
-e_{1}=\Delta(h)b^{\dagger}b+\delta e(h),
\end{align}
where $\Delta(h)\equiv \omega$ and
\begin{align}
\delta e(h) \equiv \frac{\Delta(h)}{2}+\delta e'=\begin{cases}
\sqrt{h(h-1)}-h+\frac{1}{2}&\text{if $h>h_{c}$}\,,\\
\sqrt{1-h^{2}}-\frac{1}{2}&\text{if $h<h_{c}$}\,.
\end{cases}
\end{align}

\subsection{Time-evolution of a driven quantum harmonic oscillator}
Let us provide the analytical solution of the time-evolution of a driven quantum harmonic oscillator (QHO). The Hamiltonian of the system is
\begin{equation}\label{eq_SM:Hamiltonian}
    \mathcal{H}_{t} = \frac{p^2}{2M} + \frac{M}{2}\omega_{t}^{2}x^{2},
\end{equation}
where $M$ is the mass of the oscillator and $\omega_{t}$ denotes a time-dependent driving function. Then, we introduce the $j$-th element $|\psi^{(j)}_t\rangle$ of the \emph{dynamical basis} that decomposes the solution $|\psi_t\rangle$ of the QHO dynamics. The wave-function $|\psi_t\rangle$ is returned by a time-dependent Schr\"{o}dinger equation. In the basis of the position operator spanned by the eigenfunctions $|x\rangle$, $|\psi^{(j)}_t\rangle$ is represented as 
\begin{equation}
|\psi^{(j)}_t\rangle = \int^{\infty}_{-\infty}\psi_{t,x}^{(j)}|x\rangle dx
\end{equation}
such that $|\psi_t\rangle=\sum_{j}c_{t}^{(j)}|\psi_{t}^{(j)}\rangle$, with $c_{t}^{(j)} \equiv \langle \psi_{t}^{(j)}|\psi_t\rangle$ complex numbers. The presence of the driving function $\omega_t$ entails a clear distinction between the elements of the dynamical basis $\{|\psi^{(j)}_t\rangle\}$ and the energy basis spanned by the instantaneous eigenstates 
\begin{equation}
|\phi_t^{(\ell)}\rangle \equiv \int^{\infty}_{-\infty}\phi_{t,x}^{(\ell)}|x\rangle dx \,,
\end{equation}
where the latter, by definition, originate from the eigendecomposition of the Hamiltonian $\mathcal{H}_t$. Hence, the space functions $\phi^{(\ell)}_{t,x}$ are equal to the elements of the \emph{adiabatic basis} associated to the QHO~\cite{dabrowski2016time}, i.e.,
\begin{equation}\label{eq:adiabatic_basis}
    \phi^{(\ell)}_{t,x} \equiv \frac{1}{\sqrt{2^{\ell}\ell!}}\left(\frac{\omega_{t}}{\pi}\right)^{\frac{1}{4}}e^{-\omega_{t}\frac{x^{2}}{2}}H_{\ell}\left(x\sqrt{\omega_{t}}\right)
\end{equation}
with $H_{\ell}(y)$ denoting the $\ell$-th Hermite polynomial of $y$. Only in the adiabatic limit, i.e., \emph{slow driving} (slow with respect to the energy scale of the QHO), the dynamical and adiabatic bases coincide.

As discussed in the main text, irreversible work during the drive of the QHO arises from the non-trivial overlaps $\langle \phi_{t'}^{(\ell)}|\psi_{t'}\rangle$ among the dynamical and adiabatic bases, with $\ell \in \{n,m\}$ and $t'\in[t_0,\tau]$. Hence, we are going to show the formal expression of $\langle \phi_{t'}^{(\ell)}|\psi_{t'}\rangle$. In doing this, we assume to initialize the QHO in the ground state $|\phi_{t_0}^{(0)}\rangle$ of $\mathcal{H}_{t_0}$, namely $|\psi_{t_0}\rangle = |\psi_{t_0}^{(0)}\rangle = |\phi_{t_0}^{(0)}\rangle$, corresponding to the lower-energy element of the adiabatic basis. At $t=t_0$ the overlap $\langle \phi_{t_0}^{(n)}|\psi_{t_0}\rangle$ in the position representation is 
\begin{equation}
\langle\phi_{t_0}^{(n)}|\psi_{t_0}\rangle =
\int^{\infty}_{-\infty}\phi_{t_0,x}^{(n)\ast}\psi_{t_0,x}^{(0)}dx \,,
\end{equation}
where we made use of the identity $\mathbb{I} \equiv \int^{\infty}_{-\infty} |x\rangle\!\langle x|dx$. Instead, in order to express the overlap $\langle\phi_{t'}^{(\ell)}|\psi_{t'}\rangle$ at the generic time $t'$, we have to consider the dynamics of the driven QHO that can be determined exactly~\cite{lewis1967classical,lewis1969exact,dabrowski2016time,defenu2021quantum}. In the representation of the coordinate $x$, the dynamical state of the QHO can be generally expressed as $\psi_{t,x} = \sum_{j}\alpha_{j}\,\psi_{t,x}^{(j)}$, where $\alpha_{j}$ are time-independent coefficients (complex numbers) depending on the choice of the initial state, and $\psi_{t,x}^{(j)}$ are the space elements of the QHO \emph{dynamical eigenstates}. The latter are identically equal to  
\begin{equation}
    \psi_{t,x}^{(j)} \equiv \frac{1}{\sqrt{2^{j}j!}}\left(\frac{1}{2\pi\xi_{t}^{2}}\right)^{\frac{1}{4}}e^{-\Omega_{t}\frac{x^{2}}{2}}H_{j}\left(\frac{x}{\sqrt{2}\xi_{t}}\right)e^{-i(j + \frac{1}{2})\lambda_{t}}
\end{equation}
with $\xi_{t}$ denoting the \emph{effective width} of the evolved eingestate, while 
\begin{equation}
\lambda_{t} \equiv \int_{0}^{t} \frac{1}{2\xi_{\nu}^{2}}d\nu
\end{equation}
is the \emph{total phase} accumulated till the time instant $t$. The effective width $\xi_{t}$ can be obtained by solving the \emph{Ermakov equation}
\begin{equation}\label{app_eq:EME}
    \ddot{\xi}_{t} + \omega_{t}\xi_{t} = \frac{1}{4\xi_{t}^{3}}\,,
\end{equation}
from which one gets
\begin{equation}
    \Omega_{t} = -i\frac{\dot{\xi}_{t}}{\xi_{t}}+\frac{1}{2\xi_{t}^{2}}\,.
\end{equation}
In particular, by initializing the QHO in $|\phi_{t_0}^{(0)}\rangle$, in the identity $\psi_{t,x} = \sum_{j}\alpha_{j}\,\psi_{t,x}^{(j)}$ we are allowed to replace $a_j$ with the Dirac delta $\delta(j-0)$. This entails that the QHO dynamics remains confined in the ground state but with a non-trivial time-dependence, i.e., the dynamical state $\psi_{t,x}$ associated to the position $|x\rangle$ is returned by  
\begin{equation}\label{eq:lth_dyn_state}
    \psi_{t,x} = \left(\frac{1}{2\pi\xi_{t}^{2}}\right)^{\frac{1}{4}}e^{-\Omega_{t}\frac{x^{2}}{2}}H_{0}\left(\frac{x}{\sqrt{2}\xi_{t}}\right)e^{-\frac{i}{2}\lambda_{t}}\,.
\end{equation}

\subsection{Internal energy distribution}

\subsubsection{Formal expression}

The internal energy $\Delta E_t$ is a stochastic process, whose fluctuations originate according to a specific probability distribution that depends on \textit{(i)} what measurement protocol is adopted to evaluate the energy statistics; \textit{(ii)} the possible presence of semi-classical noise terms; \textit{(iii)} the interaction with an external environment. Thus, one has to characterize its probability distribution ${\rm P}(\Delta E_t)$. In doing this, we resort to the end-point measurement (EPM) scheme, recently proposed in Refs.~\cite{GherardiniArXiv2020,HernandezGomezArxivEntropy2022}. This choice is dictated by the operational simplicity of this measurement procedure and from the observation that, under the working hypothesis of initializing the QHO in the ground state of $\mathcal{H}_{t_0}$, one obtains the same distribution of $\Delta E_t$ by applying the two-point measurement scheme~\cite{TalknerPRE2007,CampisiRMP2011}. Hence, in accordance with the EPM scheme, one has that
\begin{equation}\label{eq:energy_distribution}
    {\rm P}(\Delta E_t) = \sum_{n,m}p(E_{t_0}^{(n)})p(E_{t}^{(m)})~\delta\big(\Delta E_t - E^{(m)}_{t} + E^{(n)}_{t_0}\big),
\end{equation}
where $\delta(\cdot)$ denotes the Dirac delta and $p(E_{t_0})$, $p(E_{t})$ are the probabilities to measure one of the QHO energies at times $t_0$ and $t$ respectively.

Let us now provide the expression of the initial energy probabilities $p(E_{t_0}^{(n)}) \equiv \langle \phi_{t_0}^{(n)} |\rho_{t_0}|\phi_{t_0}^{(n)}\rangle$ with $\rho_{t_0}$ denoting the initial density operator of the QHO. In the position representation, $p(E_{t_0}^{(n)})$ is given by
\begin{equation}\label{eq:p_E0_n}
    p(E_{t_0}^{(n)}) 
    =\int^{\infty}_{-\infty}\int^{\infty}_{-\infty}\phi_{t_0,x'}^{(n)\ast}\rho_{t_0}(x',x)\,\phi_{t_0,x}^{(n)}dx\,dx'
\end{equation}
with $\rho_{t_0}(x,x') \equiv \langle x |\rho_{t_0}|x'\rangle$. Hence, since initializing the QHO in the ground state of $\mathcal{H}_{t_0}$ implies that $\rho_{t_0}=|\phi_{t_0}^{(0)}\rangle\!\langle\phi_{t_0}^{(0)}|$, one gets 
\begin{equation}\label{eq:p_E0n}
p(E_{t_0}^{(n)})=\left|\int^{\infty}_{-\infty}\phi_{t_0,x}^{(0)\ast}\phi_{t_0,x}^{(n)}\,dx\right|^{2} = \delta(n-0)\,.
\end{equation}

Then, we discuss the final energy probabilities $p(E_{t}^{(m)})\equiv\langle\phi_{t}^{(m)}|\rho_{t}|\phi_{t}^{(m)}\rangle$. We thus introduce the evolved density operator $\rho_{t} = U_{t}\rho_{t_0}U_{t}^{\dagger}$, where $U_{t} \equiv \mathcal{T}{\rm exp}(-(i/\hbar)\int^{t}_{0}\mathcal{H}_{\xi}d\xi)$ is the time-dependent propagator of the QHO dynamics with $\mathcal{T}$ denoting the time-ordering operator. Hence, one finds
\begin{equation}
    p(E_{t}^{(m)}) 
    \int^{\infty}_{-\infty}\int^{\infty}_{-\infty}\phi_{t,x'}^{(m)\ast}\rho_{t}(x',x)\,\phi_{t,x}^{(m)}dx\,dx'.
\end{equation}
In the physical context we are analyzing, the evolved density operator is the outer product of the wave-function $|\psi_{t}\rangle = U_{t}|\psi_{t_0}^{(0)}\rangle$ with $|\psi_{t_0}^{(0)}\rangle = |\phi_{t_0}^{(0)}\rangle$. This implies that $\rho_{t}(x',x) = \psi_{t,x'}\,\psi_{t,x}^{\ast}$ such that
\begin{equation}\label{eq:p_Etm}
    p(E_{t}^{(m)}) 
    = \left|\int^{\infty}_{-\infty}\psi_{t,x}^{\ast}\,\phi_{t,x}^{(m)}\,dx\right|^{2}.
\end{equation}
As a result, by looking at Eqs.~(\ref{eq:p_E0n}) and (\ref{eq:p_Etm}), the instantaneous energy probabilities $p(E_{t_0})$ and $p(E_{t})$ are returned by the squared modulus of the overlap, integrated over the space coordinates, between the eingenstates $|\phi_{t}\rangle$ of $\mathcal{H}_t$ and the solution $|\psi_t\rangle$ of the quantum dynamics, respectively. Operationally, the derivation pf $p(E_{t_0})$ and $p(E_{t})$ relies on computing at any time $t$ the product $\psi_{t,x}^{\ast}\phi^{(\ell)}_{t,x}$ for $\ell \in \{n,m\}$. 

\subsubsection{Analytical derivation}

In order to compute the joint probabilities associated to the internal energy $\Delta E_t$, we take into account the analytical expression of $\psi_{t,x}^{(j)}$, i.e.,
\begin{equation}
    \psi_{t,x}^{(j)} = \psi_{t,x}^{(0)} \equiv \left(\frac{1}{2\pi\xi_{t}^{2}}\right)^{\frac{1}{4}}e^{-\frac{1}{2}\left(\Omega_{t}x^{2}+i\lambda_{t}\right)}
\end{equation}
that, we recall, is obtained by initializing the QHO in the ground state $|\phi_{t_0}^{(0)}\rangle$ of $\mathcal{H}_{t_0}$. In this way, by employing the generating function for Hermite polynomials to express the $u$-th Hermite polynomial $H_{u}(y)$, namely
\begin{equation}
    H_{u}(y) \equiv \lim_{z \rightarrow 0}\frac{d^{u}}{dz^{u}}e^{2yz - z^2}\,,
\end{equation}
the following expression for the integral $\int^{\infty}_{-\infty}\psi_{t,x}^{\ast}\,\phi^{(m)}_{t,x}\,dx$ (over the space coordinates) is obtained:
\begin{equation}\label{app:overlap1_joint}
    \int^{\infty}_{-\infty}\psi_{t,x}^{\ast}\,\phi^{(m)}_{t,x}\,dx = e^{i\frac{\lambda_t}{2}}\left(\frac{\sqrt{\omega_t}}{m!\,2^{m+\frac{1}{2}}|\xi_t|}\right)^{\frac{1}{2}}\lim_{z\rightarrow 0}\frac{d^{m}}{dz^{m}}\int^{\infty}_{-\infty}e^{-\beta_{1}x^{2}+\beta_{2}x+\beta_{3}}\,dx
\end{equation}
where $\phi^{(m)}_{t,x}$ is provided by Eq.~(\ref{eq:adiabatic_basis}) upon substituting $k$ with $m$, and 
\begin{equation}
    \beta_1 \equiv \frac{\Omega_t + \omega_t}{2}\,;\,\,\,\,\,\beta_2 \equiv 2\sqrt{\omega_t}z\,;\,\,\,\,\,\beta_3 \equiv - z^{2}\,.
\end{equation}
The integral $\int^{\infty}_{-\infty}e^{-\beta_{1}x^{2}+\beta_{2}x+\beta_{3}}\,dx$ is the one of a Gaussian function with coefficients $\beta_1$, $\beta_2$ and $\beta_3$. We can thus resort to the analytical solution of a Gaussian integral with generic coefficients, i.e.,
\begin{equation}
    \int^{\infty}_{-\infty}e^{-\beta_{1}x^{2}+\beta_{2}x+\beta_{3}}\,dx = \sqrt{\frac{\pi}{\beta_1}}\,e^{\frac{\beta_{2}^{2}}{4\beta_{1}}+\beta_{3}}
\end{equation}
that leads us to the following simplified expression for $\int^{\infty}_{-\infty}\psi_{t,x}^{\ast}\,\phi^{(m)}_{t,x}\,dx$:
\begin{equation}
    \int^{\infty}_{-\infty}\psi_{t,x}^{\ast}\,\phi^{(m)}_{t,x}\,dx = e^{i\frac{\lambda_t}{2}}\left(\frac{\sqrt{\omega_t}}{m!\,2^{m-\frac{1}{2}}(\Omega_{t} + \omega_{t})|\xi_t|}\right)^{\frac{1}{2}}\lim_{z\rightarrow 0}\frac{d^{m}}{dz^{m}}e^{-\gamma z^2},
\end{equation}
where
\begin{equation}
    \gamma \equiv \frac{\Omega_t - \omega_t}{\Omega_t + \omega_t}\,.
\end{equation}
Accordingly, by denoting $\sqrt{\gamma}z \equiv z'$, such that $dz = dz'/\sqrt{\gamma}$, one finds: 
\begin{equation}
    \lim_{z\rightarrow 0}\frac{d^{m}}{dz^{m}}e^{-\gamma z^2} = \lim_{z'\rightarrow 0}\gamma^{\frac{m}{2}}\frac{d^{m}}{dr^{m}}e^{-{z'}^2}
    =\lim_{z'\rightarrow 0}(-1)^{m}\gamma^{\frac{m}{2}}e^{-{z'}^2}H_{m}(z')=\lim_{z\rightarrow 0}(-1)^{m}\gamma^{\frac{m}{2}}e^{-\gamma z^2}H_{m}(\sqrt{\gamma}z)
\end{equation}
such that
\begin{equation}
   \left|\lim_{z\rightarrow 0}\frac{d^{m}}{dz^{m}}e^{-\gamma z^2}            \right|=\left|\gamma^{\frac{m}{2}}H_m(0)\right| = \begin{cases}
    \displaystyle{\left|\frac{\Omega_t - \omega_t}{\Omega_t + \omega_t}\right|^{\frac{m}{2}}\frac{m!}{\frac{m}{2}!}}\,\,\,\,\text{for}\,\,\,\,m\in 2\mathbb{N} \\
    0\,\,\,\,\text{for}\,\,\,\,m\in 2\mathbb{N}+1 \,.
    \end{cases}
\end{equation}
As a result,
\begin{equation}
    p(E_{t}^{(m)}) = \frac{\sqrt{2\omega_{t}}}{|\Omega_t + \omega_t||\xi_t|}\frac{m!}{2^{m}\frac{m}{2}!\frac{m}{2}!}\left|\frac{\Omega_t - \omega_t}{\Omega_t + \omega_t}\right|^{m} = \frac{\sqrt{2\omega_{t}}}{|\Omega_t + \omega_t||\xi_t|}\frac{(m-1)!!}{m!!}\left|\frac{\Omega_t - \omega_t}{\Omega_t + \omega_t}\right|^{m}\,\,\,\,\text{for}\,\,\,\,m\in 2\mathbb{N} \,,
\end{equation}
where $2^{m}m! \equiv (2m)!!$ and $m! \equiv m!!(m-1)!!$ by definition.

Finally, since in our case-study $p(E_{t_0}^{(n)}) = \delta(n-0)$, the joint probabilities $p_{t_0,t}^{(n,m)}$ associated to the internal energy distribution are provided by the following relation:
\begin{equation}
    p_{t_0,t}^{(n,m)} = 
    p(E_{t_0}^{(n)})p(E_{t}^{(m)}) = \frac{\sqrt{2\omega_{t}}}{|\Omega_t + \omega_t||\xi_t|}\delta(n-0)\frac{(m-1)!!}{m!!}\left|\frac{\Omega_t - \omega_t}{\Omega_t + \omega_t}\right|^{m},
\end{equation}
where $n\in\mathbb{N}$ and $m\in 2\mathbb{N}$.

\subsection{Statistical moments of $\Delta E_t$ and $\nu_t$, and their relation}

In this section, we provide the analytical expression of the average internal energy $\langle\Delta E_t\rangle \equiv \langle E_{t}\rangle - \langle E_{t_0}\rangle$, and the corresponding variance ${\rm Var}(\Delta E_t) \equiv \langle\Delta E_t^2\rangle - \langle\Delta E_t\rangle^2$. Moreover, we also show their relation with the statistical moments of the number $\nu_t$ of energy excitations that originate in the evolved state of the driven QHO, as an effect of the breaking of the quantum adiabatic theorem (cfr.~Ref.~\cite{defenu2021quantum}).

At first, it is worth observing that $p(E_{t}^{(m)})$ is equal to the probability $p(\nu_{t} = m)$ of having $m$ excitations in the evolved state of the QHO. In fact, in a driven QHO the number of energy excitations is proportional to the number of energy levels of the oscillator ladder that are involved in the dynamics. Larger is the number of excitations, and higher is the energy of the levels reached by the QHO dynamics driven by the time-dependent Hamiltonian $\mathcal{H}_t$. Formally, by initializing the QHO in the ground state of $\mathcal{H}_{t_0}$, one has that $p_{t_0,t}^{(n,m)} = \delta(n-0)\,p(\nu_{t} = m)$, from which
\begin{equation}
    \langle\Delta E_t\rangle = \sum_{n,m}\Delta E_{n,m}\,p(E_{0}^{(n)})p(E_{t}^{(m)}) = \sum_{m}E_{\tau}^{(m)}p(E_{t}^{(m)}) - \sum_{n}E_{0}^{(n)}\delta(n-0) = \hbar\omega_{t}\langle \nu_{t}\rangle + \hbar\frac{(\omega_{t}-\omega_{0})}{2}\,,
\end{equation}
where 
\begin{equation}
\langle \nu_{t}\rangle \equiv \sum_{m\in 2\mathbb{N}}m\,p(\nu_{t} = m) = \frac{\xi_{t}^2}{2\omega_{t}}\left[\left(\frac{1}{2\xi^{2}_{t}} - \omega_{t}\right)^{2} + \left(\frac{\dot{\xi}_{t}}{\xi_{t}}\right)^{2}\right]
\end{equation}
is the analytical expression of the average number of energy excitations at time $t$~\cite{defenu2021quantum}.

\begin{figure}
    \centering
    \includegraphics[width=0.45\linewidth]{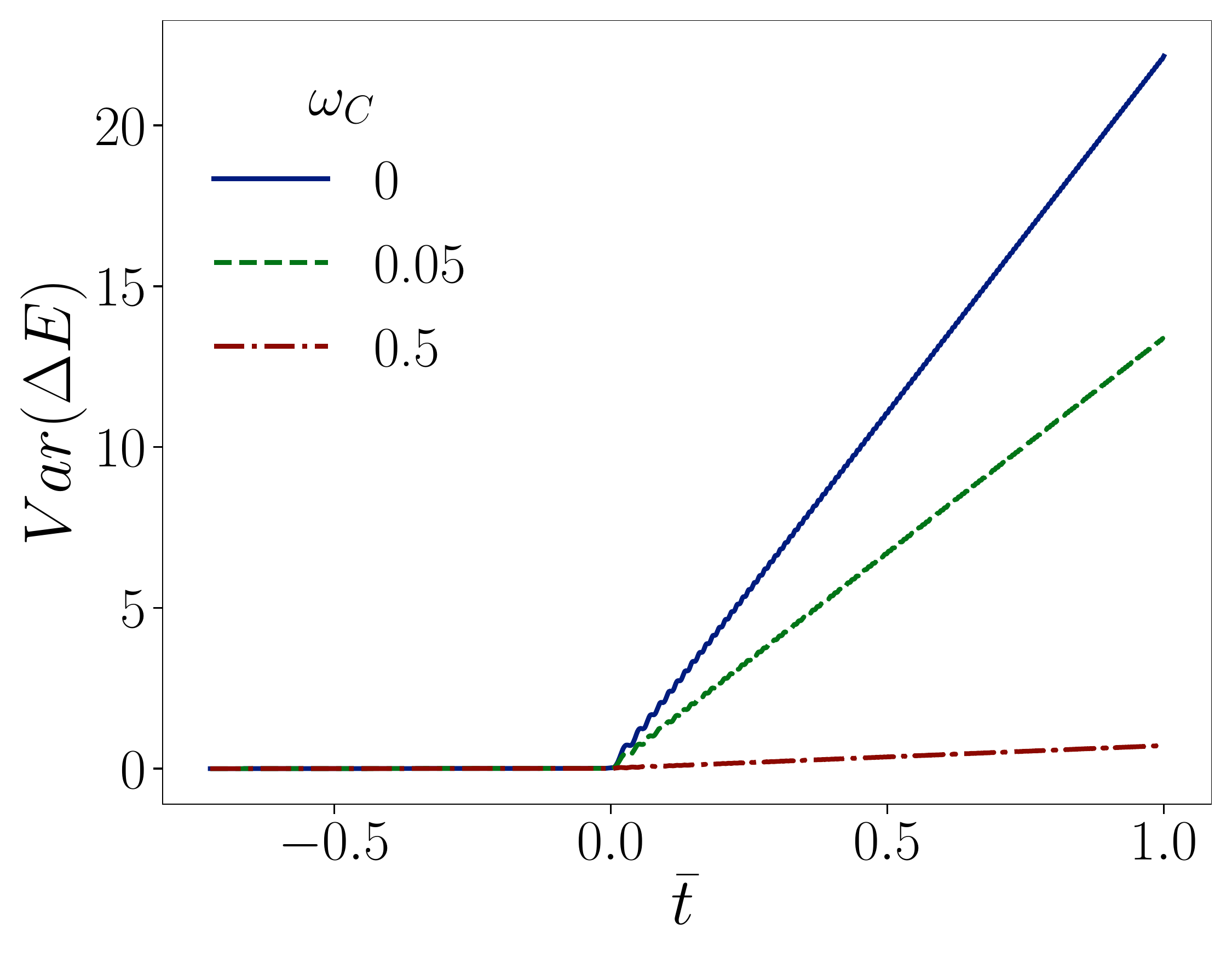}
    \caption{
    Variance of $\Delta E_t$ as a function of the normalized time $\bar{t}=t/\tau$ for different values of $\omega_C$. 
    The plotted curves are obtained by using the analytical formulas of $\langle\Delta E_t\rangle$ and ${\rm Var}(\Delta E_t)$, and the numerical solution of the Ermakov equation that provides us the values of $\xi_{t}$ and $\dot{\xi}_{t}$.
    }
    \label{fig:meanvar}
\end{figure}

Instead, to get the variance ${\rm Var}(\Delta E_t)$ of the internal energy $\Delta E_t$, we compute the second statistical moment $\langle\Delta E_t^{2}\rangle$. Thus, 
\begin{eqnarray}\label{app:second_moment}
    \langle\Delta E_t^{2}\rangle &=& \sum_{n,m}\Delta E_{n,m}^{2}\,p(E_{t_0}^{(n)})p(E_{t}^{(m)}) = \sum_{m}(E_{t}^{(m)})^{2}p(\nu_{t} = m) + (E_{t_0}^{(0)})^{2} - 2E_{0}^{(0)}\sum_{m}E_{t}^{(m)}p(\nu_{t} = m)\nonumber \\
    &=& \sum_{m}(E_{t}^{(m)})^{2}p(\nu_{t} = m) + E_{t_0}^{(0)}\left(E_{t_0}^{(0)} - 2\sum_{m}\hbar\omega_{t}\left(m+\frac{1}{2}\right)p(\nu_{t} = m)\right)\nonumber \\
    &=& \hbar^{2}\omega_{t}^{2}\left(\langle \nu_{t}^{2}\rangle + \langle \nu_{t}\rangle + \frac{1}{4}\right) - \frac{\hbar^{2}\omega_{t_0}}{4}\left(2\omega_{t}(1+2\langle \nu_{t}\rangle)-\omega_{t_0}\right)\nonumber \\
    &=& \hbar^{2}\left(\frac{\omega_{t}-\omega_{t_0}}{2}\right)^2 + \hbar^{2}\omega_{t}^{2}\langle \nu_{t}^{2}\rangle + \hbar^{2}\omega_{t}(\omega_{t}-\omega_{t_0})\langle \nu_{t}\rangle\,,
\end{eqnarray}
where $\langle\nu_{t}^{2}\rangle$ denotes the second statistical moment of the number $\nu_t$ of energy excitations. As a result, being ${\rm Var}(\Delta E_t) \equiv \langle\Delta E_t^2\rangle - \langle\Delta E_t\rangle^2$, the variance of the internal energy is finally provided by
\begin{equation}\label{app:variance}
    {\rm Var}(\Delta E_t) = \sum_{n,m}\Delta E_{n,m}^{2}\,p_{t_0,t}^{(n,m)} - \langle\Delta E_t\rangle^{2} = \hbar^{2}\omega_{t}^{2}{\rm Var}\left(\nu_{t}\right)
\end{equation}
with ${\rm Var}(\nu_{t}) \equiv \langle \nu_{t}^{2}\rangle - \langle \nu_{t}\rangle^{2}$ variance of $\nu_t$. In Fig.~\ref{fig:meanvar}, we plot the average and variance of $\Delta E_t$ as a function of the normalized time $\bar{t}=t/\tau$, with $\tau=25$ total duration of the protocol, for $\omega_C \in \{0,0.05,0.5\}$. Notice that $t$, $\tau$ and $\omega_C$ are all expressed in dimensionless units.

To conclude our derivation, we provide a way to analytically compute the statistical moments of the energy excitations number $\nu_t$. For such a purpose, it is convenient to employ the notation used in Ref.~\cite{dabrowski2016time} and thus to define the quantity
\begin{equation}
    |R_t|^{2}=\frac{\left(\frac{1}{2\xi_{t}^{2}}-\omega_{t}\right)^{2}+\frac{\dot{\xi}_{t}^{2}}{\xi_{t}^{2}}}{\left(\frac{1}{2\xi_{t}^{2}}+\omega_{t}\right)^{2}+\frac{\dot{\xi}_{t}^{2}}{\xi_{t}^{2}}}\,,
\end{equation}
so that the excitation probability $p(\nu_{t} = m)$ is equal to
\begin{equation}
    p(E_{t}^{(m)}) = \frac{(m-1)!!}{m!!}\sqrt{1-|R_t|^{2}}\,|R_t|^{m}\,.
\end{equation}
In the main text, we have outlined that this probability law is a negative binomial distribution, by showing how it can be employed to describe the statistics of topological defects formation in bosonic systems. Accordingly, in this perspective, all the statistical moments of the distribution can be easily calculated, as a function of $|R_t|^2$. Thus, e.g., for the first three moments of the excitation probability $p(\nu_{t} = m)$ one gets:
\begin{align}
    \langle \nu_{t}\rangle&=\frac{|R_t|^{2}}{1-|R_t|^{2}}\\
    \langle \nu_{t}^{2}\rangle&=\frac{|R_t|^{2}(2+|R_t|^{2})}{(1-|R_t|^{2})^{2}}\\
    \langle \nu_{t}^{3}\rangle&=\frac{|R_t|^{2}(4+10|R_t|^{2}+|R_t|^{4})}{(1-|R_t|^{2})^{3}} \,.
\end{align}

\subsection{Universality}

In all the derivations above, we have not specified a shape for the driving function $\omega_t$ with $t\in[t_0,\tau]$. We have just assumed, indeed, that $\omega_t$ is a cyclic transformation crossing the quantum critical point $\omega_C \approx 0$. Now, in order to show the universality of the distributions for both the formation of topological defects and the internal energy around the criticality, we substitute
\begin{equation}
    \omega_t = \lim_{\delta\rightarrow 1}\delta|t|^{\eta} = |t|^{\eta}\,,
\end{equation}
where $\eta$ (real number $\geq 0$) and $\delta$ are, respectively, the \emph{critical gap-scaling exponent} and the \emph{driving strength}. As proved in~\cite{defenu2021quantum}, the choice to operate in the limiting case of $\delta\rightarrow 1$ does not make us lose in generality. In fact, for a generic value of $\delta$, we can always re-scale the time-dependence of the system dynamics via the transformation $\widetilde{t}(\delta) = |\delta|^{-\frac{\eta}{1+\eta}}t$, and then use the solution obtained for $\delta = 1$. It is worth noting that this transformation is ruled by $\eta$ that is also responsible for the complete parametrization of $\psi_{t,x}$, Eq.~(\ref{eq:lth_dyn_state}), at the quantum criticality. In such points, the driving function $\omega_t$ is uniquely determined by $\xi_t$, where $\xi_{t}$ is returned by solving the Ermakov equation with initial conditions $\xi_{t_0}^{-2}/2 = \omega_{t_0}$ and $\dot{\xi}_{t_0} = 0$ that ensure that at the initial time $t=t_0$ the QHO is in the ground state (lower-energy element of the adiabatic basis). Moreover, implicitly, the initial time $t_0$ is taken sufficiently far from $t=t_c$, time at which the quantum critical point is crossed.
Specifically, as also discussed in Refs.~\cite{dabrowski2016time,defenu2021quantum}, $\xi_t$ fulfills the equation
\begin{equation}\label{eq_SM:xi_criticality}
    \xi_{t}^{2}(p) = p|t|\left(p\,\sigma_{1}^{2}+\sigma_{2}^{2}\right)\left(J_{-p}^{2}\left(\zeta_{t}(p)\right)+J_{p}^{2}\left(\zeta_{t}(p)\right)\right)
    + 2p|t|\left(p\,\sigma_{1}^{2}-\sigma_{2}^{2}\right)J_{-p}\left(\zeta_{t}(p)\right)J_{p}\left(\zeta_{t}(p)\right)
\end{equation}
with $t$ in the neighbourhood of $t_c$. In Eq.~(\ref{eq_SM:xi_criticality}), $J_{y}(x)$ are Bessel functions of the first kind, while 
\begin{equation}\label{eq_SM:sigma_before}
    \sigma_{1} = \sigma_{1}^{-} \equiv \sqrt{\frac{\pi}{2}}\frac{1}{\sqrt{p}}\frac{1}{2\cos\left(\frac{p\pi}{2}\right)};
    \,\,\sigma_{2} = \sigma_{2}^{-} \equiv \sqrt{\frac{\pi}{2}}\frac{1}{2\sin\left(\frac{p\pi}{2}\right)}
\end{equation}
for $t<t_{c}$, and
\begin{equation}\label{eq_SM:sigma_after}
    \sigma_{1} = \sigma_{1}^{+} \equiv \sqrt{\frac{\pi}{2}}\frac{1}{\sqrt{p}}\frac{1}{2\sin\left(\frac{p\pi}{2}\right)};
    \,\,\sigma_{2} = \sigma_{2}^{+} \equiv \sqrt{\frac{\pi}{2}}\frac{1}{2\cos\left(\frac{p\pi}{2}\right)}
\end{equation}
for $t\geq t_c$. Thus, Eqs.~(\ref{eq_SM:xi_criticality}), (\ref{eq_SM:sigma_before}), (\ref{eq_SM:sigma_after}) only depend on $p$ and $\zeta_t$ that are functions of the critical gap-scaling exponent $\eta$, i.e.,
\begin{equation}
    p \equiv \frac{1}{2(1+\eta)}\quad \text{and} \quad \zeta_t(p) \equiv \frac{|t|^{1+\eta}}{1+\eta} = 2p|t|^{\frac{1}{2p}} \,.
\end{equation}

\twocolumngrid
\bibliography{references.bib}

\end{document}